\documentclass[conference]{IEEEtran}
\IEEEoverridecommandlockouts

\usepackage{cite}
\usepackage{amsmath,amssymb,amsfonts}
\usepackage{algorithmic}
\usepackage{graphicx}
\usepackage{textcomp}
\usepackage{xcolor}
\usepackage{xspace}
\usepackage{tabularx}
\usepackage{booktabs}
\usepackage{fancyhdr}
\usepackage[hyphens]{url}

\def\BibTeX{{\rm B\kern-.05em{\sc i\kern-.025em b}\kern-.08em
    T\kern-.1667em\lower.7ex\hbox{E}\kern-.125emX}}

\usepackage{bm}
\usepackage{amsthm}

\newcommand{\proj}{AutoSVA\xspace}
\newcommand{\nomodules}{7\xspace}

\definecolor{darkgreen}{rgb}{0.0, 0.5, 0.0}

 \fancypagestyle{firstpage}{
 
   \fancyhead[C]{\normalsize{To Appear in \textit{Proceedings of the 58th ACM/IEEE Design Automation Conference (DAC’21)}} } 
   \fancyfoot[C]{\mycopyrightnotice}
 }

\def\mycopyrightnotice{
  {\footnotesize 978-1-6654-3274-0/21/\$31.00~\copyright2021 IEEE\hfill}
  \gdef\mycopyrightnotice{}
}

\begin{document}

\title{\proj: Democratizing Formal Verification of RTL Module Interactions}



\author{\IEEEauthorblockN{Marcelo Orenes-Vera,
Aninda Manocha, David Wentzlaff and
Margaret Martonosi}
\IEEEauthorblockA{Department of Computer Science and Electrical Engineering,
Princeton University\\
Princeton, New Jersey, USA\\
Email: \{movera, amanocha, wentzlaf, mrm\}@princeton.edu }}


\maketitle
\thispagestyle{firstpage}
\pagestyle{plain}

\begin{abstract}
Modern SoC design relies on the ability to separately verify IP blocks relative to their own specifications.
Formal verification (FV) using SystemVerilog Assertions (SVA) is an effective method to exhaustively verify blocks at unit-level.
Unfortunately, FV has a steep learning curve and requires engineering effort that discourages hardware designers from using it during RTL module development. 
We propose \proj{}, a framework to automatically generate FV testbenches that verify liveness and safety of control logic involved in module interactions.
We demonstrate \proj{}'s effectiveness and efficiency on deadlock-critical modules of widely-used open-source hardware projects.
\end{abstract}

\vspace{-1mm}
\begin{IEEEkeywords}
automatic, modular, formal, verification, SVA
\end{IEEEkeywords}
\vspace{-1mm}
\section{Introduction}

\begin{figure}[t]
    \centering
    \vspace{-2mm}
    \includegraphics[width=\columnwidth]{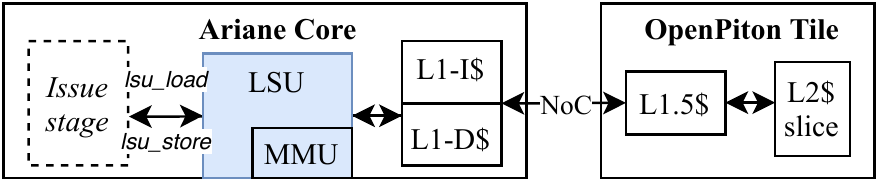}
    \vspace{-8mm}
    \caption{Heterogeneous SoCs such as OpenPiton+Ariane, involve dependencies between modules. Verifying these interactions is critical to guarantee forward progress in the system. For example, a load request to the LSU (blue) must receive a response for a memory request to eventually complete. 
    }
    \vspace{-5mm}
    \label{fig:modules}
\end{figure}

Heterogeneous SoC design is a lengthy, expensive process that necessitates verification at early stages of development to avoid late bug fixes that thwart performance or area goals~\cite{arm}.
SoC modules may be developed in various contexts and exhibit complicated interactions~\cite{byoc}. 
With the numerous dependencies that occur between them, module interface verification is necessary to prevent opportunities for livelock and deadlock.
Fig.~\ref{fig:modules} presents the Ariane core~\cite{ariane} and the cache hierarchy of the OpenPiton manycore~\cite{piton+ariane}, used as an example throughout the paper. Among the module interactions, the Load-Store Unit (LSU) is critical for the forward progress of the system.

SystemVerilog Assertions (SVA)~\cite{sva_spec} is often used for RTL verification because it is a powerful language for defining a design's properties and specifying temporal dependencies.
SVA properties can be checked through both test-driven simulation and Formal Verification (FV) in order to reveal bugs. However, only FV tools can exhaustively test a given Design-Under-Test (DUT) and consequently are most suitable for verifying forward progress~\cite{jg_user,symbiotic}. 
Unfortunately, these tools present a steep learning curve and require significant engineering effort to set up a \textit{useful} FV testbench, i.e. writing appropriate properties and specification constraints.
\textit{This upfront knowledge and effort discourages hardware designers from using FV}~\cite{formal_book}.
Some tools generate SVA from a higher abstraction layer~\cite{ilang, rtlcheck}, but creating a high-level model and mapping it to RTL signals is still cumbersome. 
These tools also do not cover important properties like forward progress.

With the goal of making FV \textit{agile} and \textit{widely used} among hardware designers, we make two \textbf{key observations}: 
(1)~Although interactions between RTL modules may take place via different mechanisms, a common design pattern across many of them is \textit{request and response}.
We advocate for automatic support of FV for this pattern.
(2)~Capturing the request/response abstraction in a model allows for \textit{automated reasoning} about RTL interfaces and their expected interactions.
This work proposes a language centered around a \textit{transaction} model.
This model's applicability is not limited to modules with explicit requests/responses; it can express other interface mechanisms, e.g. pipeline stages that receive requests from a previous stage and send them to the next stage.

\textbf{Approach:} Given our observations, this work proposes \proj{}, a framework to automatically generate Formal verification Testbenches (FT) for a given DUT. The designer of the DUT only needs to identify relevant transactions and annotate them in the module interface using a simple language. The framework then generates properties that verify the transactions are \textit{well-formed} and make \textit{forward-progress}: they satisfy \textit {liveness} (every request is eventually followed by a response) and \textit{safety} (expectations for attributes of the response). 
Through its automated reasoning, \proj creates the necessary scaffolding code to express these properties and tool-specific commands to drive the FV process, alleviating the hardware designer from significant engineering effort and democratizing the use of FV for verifying forward progress.

FTs generated by \proj can then be supplied to a FV tool, e.g. JasperGold~\cite{jg_user} or the open-source SymbiYosys tool~\cite{symbiotic}. 
\proj thereby provides a frontend for automatic FV of an important subset of the correctness problem---ensuring RTL modules' interface expectations. 

\textbf{Our main contributions are:}

\vspace{-0.5mm}
\begin{itemize}
    \item A language that creates a unified transaction abstraction to denote RTL interface interactions and dependencies. This enables automatic reasoning about RTL properties.
    \item An automated procedure to generate FTs
    that express liveness properties about transaction temporal dependencies and safety properties about control-logic attributes. 
    \item Demonstration of \proj{}'s effectiveness on \nomodules control-critical RTL modules of the widely-used open-source projects Ariane and OpenPiton~\cite{ariane,piton+ariane}.
    As one example, within 1 hour, \proj generated a FT for Ariane's MMU, discovered a bug, and verified the bug-fix.
\end{itemize}



\section{Motivating Example}

\begin{figure}[t]
    \centering
    \vspace{-4mm}
    \hspace*{-2mm}
    \includegraphics[width=\columnwidth]{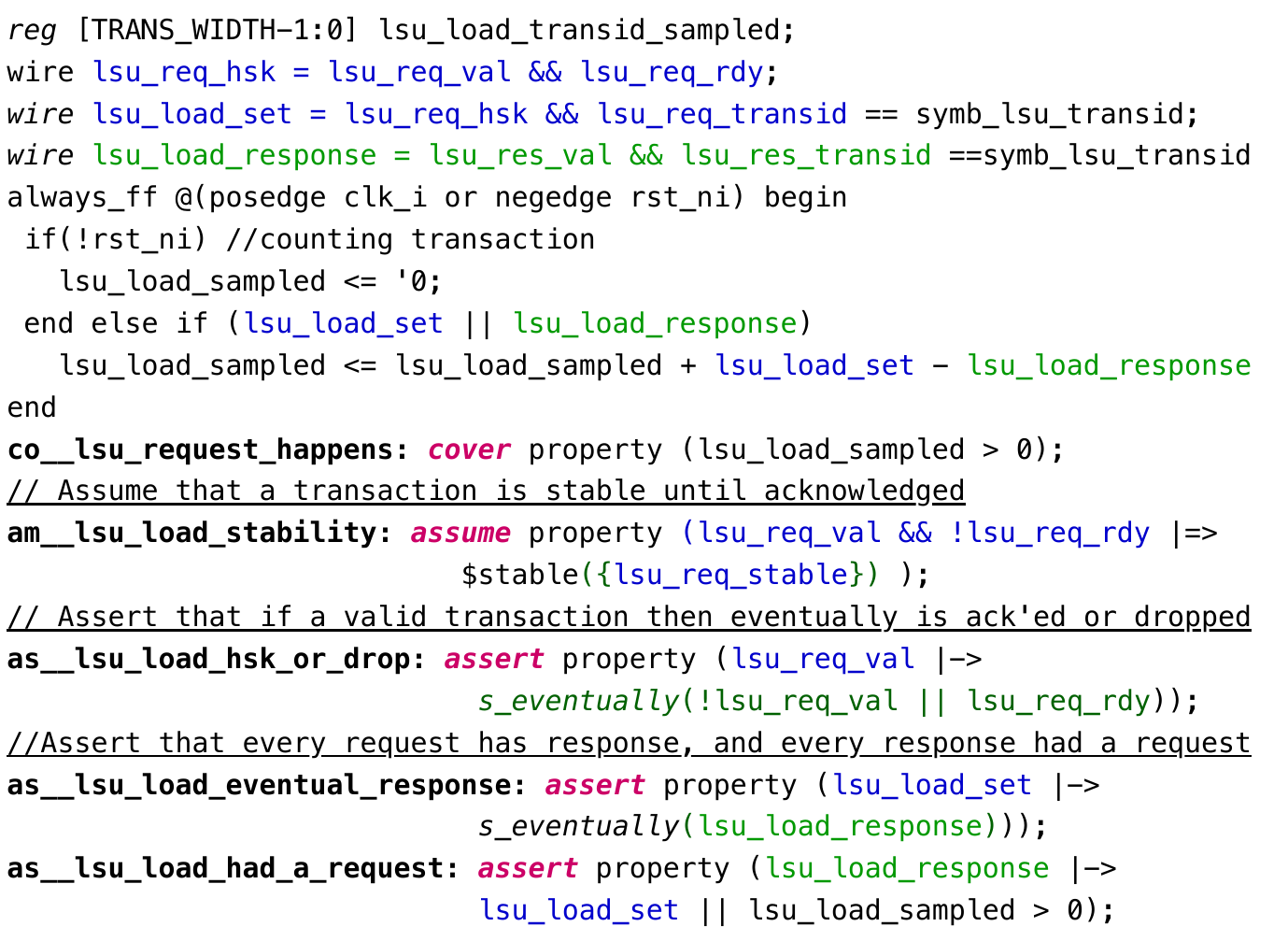}
    \vspace{-5mm}
    \caption{To verify the load interface of the LSU, a hardware designer would need to write many SVA properties and auxiliary code. \proj automatically generates all modeling, removing the burden from the designer.}
    \label{fig:sva_example}
    \vspace{-4mm}
\end{figure}

SVA is SystemVerilog's formal specification language~\cite{sva_spec}, and offers a mature approach for verifying RTL.
It can express Linear Temporal Logic (LTL) formulas over interface signals to build properties about module interactions. LTL specifies temporal relations, which fall into two major classes: safety and liveness properties. \textit{Safety} properties specify that “nothing bad will happen”, e.g. a response must have had a request; while \textit{liveness} specify that “something good will happen”, e.g. a request is eventually acknowledged.

Fig.~\ref{fig:sva_example} presents a subset of the modeling and properties that are necessary to verify forward progress for the load-store unit (LSU) in Ariane (depicted in Fig.~\ref{fig:bug1}).
For example, \textit{lsu\_load\_eventual\_response} is a liveness property to  check that any load request eventually receives a response with the same transaction ID.
Verifying expectations about module interfaces goes beyond writing properties; it requires code to sample transactions and symbolic variables to track them.
\proj automatically generates all of this necessary code. 

Properties in SVA can use one of three directives: \textit{assert}, \textit{assume} and \textit{cover}. 
Assumptions have different meanings based on how input stimuli are generated. In RTL simulation, inputs are driven either by manual or random tests, and thus \textit{assume} has the same meaning as \textit{assert}, i.e they check that the property holds. 
Conversely, FV tools treat inputs as Boolean variables, and \textit{assumptions} constrain the state space exploration by preventing some behaviors, while \textit{assertions} check that properties hold on the explored paths.
FV tools then use a variety of solver engines~\cite{jg_engine} based on formal methods, such as model checking, which uses SAT (satisfiability)~\cite{smc_no_bdds} or BDD (binary decision diagrams)~\cite{smc_mcmillan}, to exhaustively search for property violations.
FV search may result in a counterexample (CEX) that highlights the violation of a property, or proof that properties hold, i.e. the solver converges and finds no CEXs. 

The underlying dynamics of FV and SVA make it difficult to intuitively understand the consequences of various properties expressed, such as the behavior of symbolic variables, e.g.\ \textit{symb\_lsu\_transid} in Fig.~\ref{fig:sva_example}, which allow the tracking of indices with a single assertion.
Furthermore, subtle mistakes in assumptions, e.g.\ using the $|$$-$$>$ implication symbol in the \textit{lsu\_stability} assumption, can \textit{over-constrain} the state space and end up proving vacuity. Manually inserting assertions can be cumbersome and error-prone for a hardware designer, and particularly frustrating when CEXs appear due to illegal inputs of not yet modeled interfaces~\cite{cummings2016}.
Thus, \proj automatically models and expresses the expected behavior for module transactions.

\begin{figure}[t]
    \centering
    \vspace{-6mm}
    \hspace*{+1mm}
    \includegraphics[width=0.97\columnwidth]{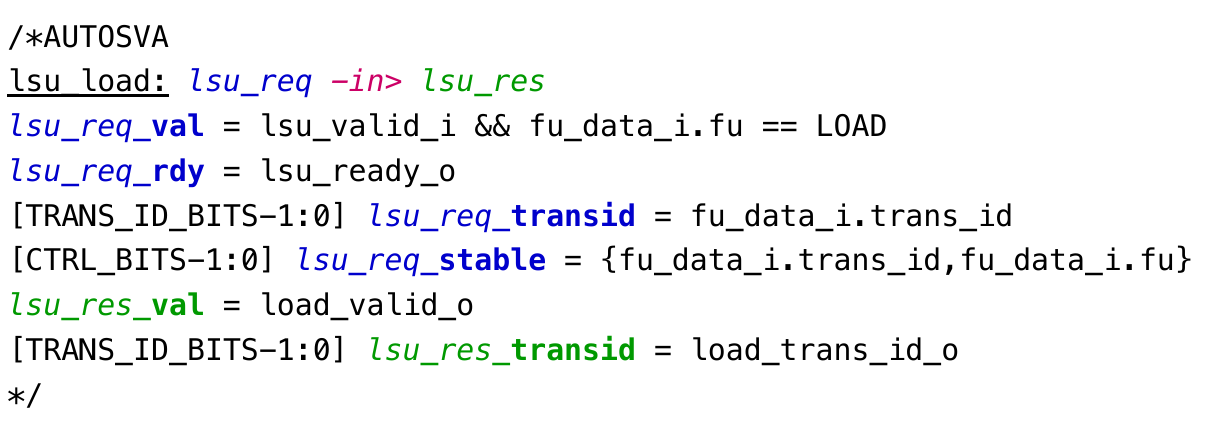}
    \vspace{-6mm}
    \caption{The LSU designer only needs to annotate the RTL interface using \proj{}'s language to generate a FT (containing among other things the properties and modeling code shown in Fig.~\ref{fig:sva_example}). The first line describes a relation between a request (italic blue) and a response (italic green) interface; the remaining lines map RTL interface signals to transaction attributes (bold).} 
    \label{fig:auto_sva_example}
    \vspace{-4mm}
\end{figure}

Hardware designers can employ SVA properties for Test-Driven-Development (TDD), where CEXs help to refine the design~\cite{sutherland2015}. Moreover, this can be applied at early stages of RTL module development by using unit-level FV~\cite{tdd_formal}. However, FV's steep learning curve and necessary engineering effort preclude designers from using it in practice~\cite{formal_book}.
\textbf{\proj democratizes FV for hardware designers and makes TDD practical by automating a key component of the FV problem: \textit{liveness and safety of module interfaces}.} 
Instead of aiming to support functional FV, which is very implementation-dependent, \proj focuses on verifying that modules interact through \textit{well-formed transactions}. This verification entails checking certain attributes over the control logic involved in transactions and mapping them to properties that ensure that every module makes progress (does not hang). 

Fig.~\ref{fig:auto_sva_example} presents an example of the simple usage of \proj{}'s language. 
These annotations unleash automated reasoning to generate the modeling and properties (shown in Fig.~\ref{fig:sva_example}) surrounding liveness and safety for the Ariane core's LSU .  Section~\ref{sec:approach} explains the syntax of the \proj language and semantics of each annotation. Section~\ref{sec:evaluation} shows how these annotations can be applied to different interface styles.




\vspace{-1mm}
\section{The \proj{} Framework \label{sec:approach}}



\proj focuses on verifying liveness, and its well-formed transactions allow it to utilize a common abstraction from RTL interfaces that avoids the complexity of specific module implementations. 
By capturing a common design-pattern, \proj{} can automatically generate \textit{useful} Formal Testbenches (FT). We denote a FT as \textit{useful} when it (1)~has sufficient module interface modeling to avoid spurious CEXs and capture relevant CEXs which lead to uncovering bugs, and (2)~does not miss legal scenarios due to over-constraining assumptions. Moreover, \proj reduces the state-explosion scalability problem because it deliberately focuses on control logic and FV tools can be instructed to automatically ignore datapaths.

\begin{figure}[t]
    \vspace{-2mm}
    \centering
    \includegraphics[width=0.9\columnwidth]{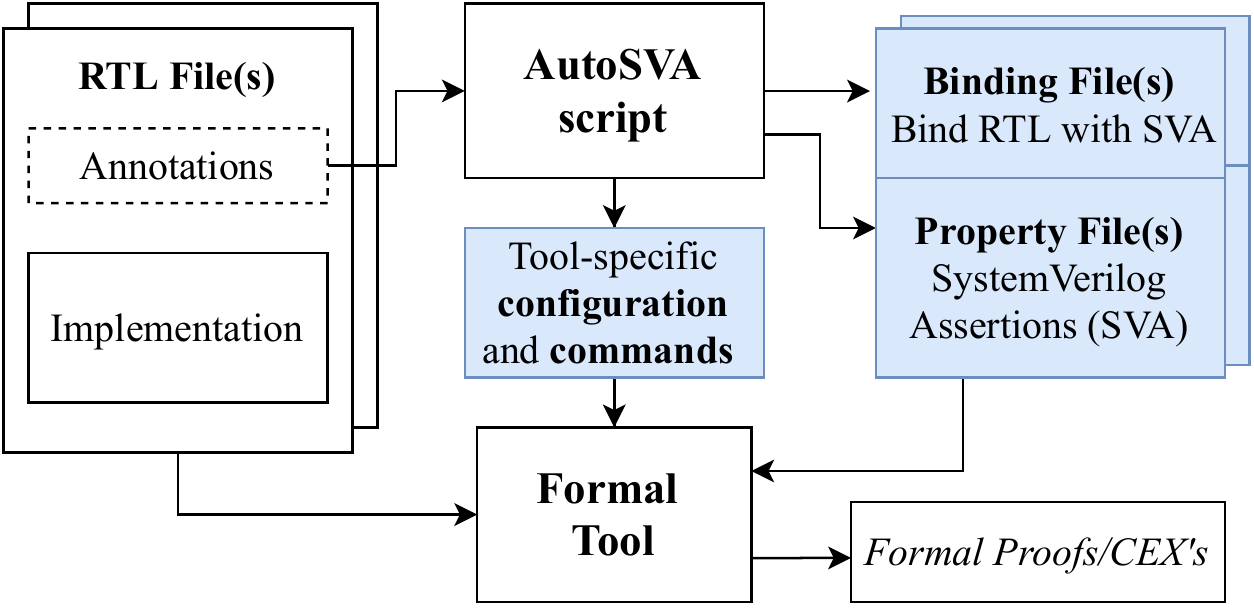}
    \vspace{-1.5mm}
    \caption{\proj{} is an agile framework for FV of RTL using SVA. The files that define the FT are denoted in blue. Dotted lines indicate designer input.}
    \label{fig:autosva_overview}
    \vspace{-2mm}
\end{figure}

Fig.~\ref{fig:autosva_overview} presents an overview of \proj{}'s verification process. \proj{} takes as input the interface declaration section of the RTL module acting as the DUT.
The interfaces should be annotated using \proj{}'s language for interface abstraction (defined at Section \ref{sec:abstraction}). 
Once the abstraction is defined for a DUT, \proj{} generates the FT that includes a property file describing the properties to verify, all necessary modeling about RTL blocks external to the DUT, and a binding file to connect the properties to signals in the DUT. 

Based on the FV tool to target, \proj{} generates configuration and command files.
\proj{} currently supports JasperGold~\cite{jg_user} and SymbiYosys~\cite{symbiotic}.
Once the properties, binding and tool-specific files are generated, \proj{} invokes the FV tool to start the verification process. This returns either property proofs or CEXs that highlight possible bugs in the RTL. 
A hardware designer can then quickly set up a FT and locate bugs by using \proj as a frontend for FV tools.

\subsection{\proj Language to Express Transactions \label{sec:abstraction}}

\proj{}'s transaction abstraction involves two events connected with an implication relation. From the DUT's perspective there are two types of transactions: (1)~\textit{incoming} transactions describe when a DUT receives a request and is responsible for eventually triggering a well-formed response or another request, and (2)~\textit{outgoing} transactions describe when a DUT triggers a request that eventually must receive a response.

The two events in a transaction are associated with RTL \textit{interfaces}, which are the connection points of RTL modules. For example, incoming transactions can map a cache lookup interface to define a liveness condition that the cache lookup should eventually have a response, and to define a safety condition that this response must satisfy certain properties, e.g. maintain the same transaction ID the request had.

\proj maps transaction events to interfaces through annotations expressed in its language. These language annotations are written as Verilog comments on the interface declaration section of an RTL file to identify module interfaces that participate in transactions. To distinguish these annotations from regular code comments, \proj requires annotations to be preceded with an \textit{AUTOSVA} macro, or be contained within a multi-line comment region that starts with it.

\begin{scriptsize}
\begin{table}[t]
\vspace{-5mm}
\centering
\caption{THe \proj{} language. Constants are written in lowercase and syntax in uppercase. STR and ASSIGN are Verilog's syntax for strings and assignments. }
\vspace{-2mm}
\begin{tabularx}{\columnwidth}{l}
\toprule
\textit{TRANSACTION ::= TNAME: {\color{purple} \textbf{RELATION}} \textbf{ATTRIB}} \\
\textit{\color{purple} \textbf{RELATION} ::= {\color{blue} \textbf{P}} $-$in$>$ {\color{darkgreen} \textbf{Q}} $|$ {\color{blue} \textbf{P}} $-$out$>$ {\color{darkgreen} \textbf{Q}}} \\
\textit{\textbf{ATTRIB} ::= \textbf{ATTRIB}, \textbf{ATTRIB} $|$ SIG = ASSIGN $|$ input SIG $|$ output SIG} \\
\textit{SIG ::= [STR:0] FIELD $|$ STR FIELD} \\
\textit{FIELD ::= {\color{blue} \textbf{P}}\_\textbf{SUFFIX} $|$ {\color{darkgreen} \textbf{Q}}\_\textbf{SUFFIX}} \\
\textit{\textbf{SUFFIX} ::= val $|$ ack $|$ transid $|$ transid\_unique $|$ active $|$ stable $|$ data} \\
\textit{TNAME, {\color{blue} \textbf{P}}, {\color{darkgreen} \textbf{Q}} ::= STR} \\
\bottomrule
  \end{tabularx}
  \vspace{-1mm}
  \label{table:language}
\end{table}
\end{scriptsize}

Table \ref{table:language} presents the formalization of the \proj{} language.
$P$ and $Q$ represent two interfaces which have a temporal implication relation, which is either incoming ``\textit{$-$in$>$}'' or outgoing ``\textit{$-$out$>$}'' from the DUT's perspective, and share a transaction named \textit{TNAME}. Multiple transactions can be defined with unique names.
\textit{ATTRIB} definitions map interface signals to transaction attributes.
Each definition must be placed on a separate line in the RTL, i.e. distinct line number, and must be prefixed with the interface name. 

\textit{Implicit definitions} are native interface signal declarations (preceded by input/output signals) that are already defined in the RTL design. If they follow the \textit{FIELD} naming convention, \proj can automatically identify these fields without annotations, which is especially useful for early-stage RTL verification. \proj{}'s parser ignores signal declarations that do not match $P$ or $Q$ prefixes and the language's legal suffixes.

\textit{Explicit definitions} define new signals to extract transaction attributes that are not explicitly defined with interface signals.
These are useful for renaming signals that do not match \proj{}'s language, extracting fields within structs, and defining attributes based on multiple interface signals. Fig.~\ref{fig:annotations2} presents examples of these definitions for a few modules.

\subsection{Property Generation Based on Transaction Attributes \label{sec:properties}}
 
\begin{scriptsize}
\begin{table}[t]
    \centering
    \vspace{-2.5mm}
    \caption{Properties generated for each transaction attribute.}
    \vspace{-2mm}
    \begin{tabularx}{\columnwidth}{l l}
        \toprule
        \textbf{Attribute} & \textbf{Properties generated} \\
        \midrule
        \textit{val$^{*}$} & If \textit{P} is valid, then eventually \textit{Q} will be valid and\\
            & for each \textit{Q} valid, there is a \textit{P} valid \\
        \midrule
        \textit{ack$^{*}$} & If \textit{P} is valid, eventually \textit{P} is ack'ed or \\
                & \textit{P} is dropped (if its \textit{stable} signal is not defined) \\
        \midrule
        \textit{stable} & If \textit{P} is valid and not ack'ed, then it is \textit{stable} next cycle \\
        \midrule
        \textit{active} & This signal is asserted while transaction is ongoing \\
        \midrule
        \textit{transid$^{*}$} & Each \textit{Q} will have the same transaction ID as \textit{P} \\
        \midrule
        \textit{transid\_unique} & There can only be 1 ongoing transaction per ID \\
        \midrule
        \textit{data$^{*}$} & Each \textit{Q} will have the same data as \textit{P} \\
        \bottomrule
    \end{tabularx}
\label{table:attributes}
\vspace{-3mm}
\end{table}
\end{scriptsize}

\proj generates properties based on how transactions are defined, as more attributes indicate more characteristics to verify.
Table~\ref{table:attributes} presents the properties that result from the presence of each attribute.
\proj{} does not require all possible transaction attributes to be defined in order to generate meaningful properties. For example, an implication relation between $P$ and $Q$ with just the \textit{val} attribute defined indicates the two interfaces communicate and thus a liveness property is generated for the transaction. The absence of an \textit{ack} signal indicates the request/response is always accepted.

Some of the properties expressed in Table~\ref{table:attributes} cannot be expressed in SVA directly, and thus \proj generates all necessary auxiliary Verilog code. For example, verifying that every response followed a previous request requires counting the number of ongoing transactions (done with registers in Fig~\ref{fig:sva_example}). 
The \textit{transid} attribute allows tracking transactions to reason about other attributes, such as \textit{data}, which is used to verify data integrity. This is important for interface fields which are immutable between $P$ and $Q$, e.g. data in a queue or address in a memory request.

Attributes marked with \textit{*} at Table~\ref{table:attributes} generate properties that are asserted when the transaction is \textit{incoming} and assumed when \textit{outgoing}.
E.g., for the \textit{val} attribute, the word "eventually" indicates liveness when the DUT is expected to respond and fairness when it is waiting for a response. For attributes \textit{stable} and \textit{transid\_unique}, the opposite holds; properties are assumed on incoming and asserted on outgoing transactions. The attribute \textit{active} is always asserted when defined.

\textit{Submodule Properties:}
When the DUT has a submodule whose inputs are driven by actual logic, it is worthwhile to ensure that assumptions about these inputs hold.
\proj assumptions can be converted into assertions by changing the value of the \textit{ASSERT\_INPUTS} parameter. Submodule properties can be linked to the parent's through \proj{}'s parameters: "\textit{-AM}" includes the properties when the submodule was the DUT (assumptions over outgoing requests) and "\textit{-AS}" converts all assumptions into assertions.

\textit{End-to-End Properties:} SVA allows writing properties that use internal RTL logic (not visible at the interface). While this is often necessary for full functional verification, it causes properties to depend on RTL implementation details.
To overcome this, \proj writes end-to-end properties which solely describe interface signals, but cover the whole path from input to output interface. End-to-end properties are implementation-agnostic, and thus can be automatically generated pre-RTL, making \proj a great framework for Test-Driven-Development (TDD).

\textit{Property Reuse:} In addition to FV, \proj property files can be utilized in a simulation testbench to ensure that assumptions hold during system-level testing.
Although many RTL simulation tools do not support liveness properties, all control-safety properties and X-propagation assertions can be checked during simulation.
\proj generates X-propagation assertions, which check that when the \textit{val} signal of an interface is asserted, none of the other attributes have value \textit{X} (concurrently 0 and 1). 
Because formal tools do not consider X's and instead assign arbitrary values of 0 or 1, these assertions are only checked during simulation (under a \textit{XPROP} macro).

\subsection{\proj{} Implementation and Process Steps}

\begin{figure}[t]
    \centering
    \includegraphics[width=\columnwidth]{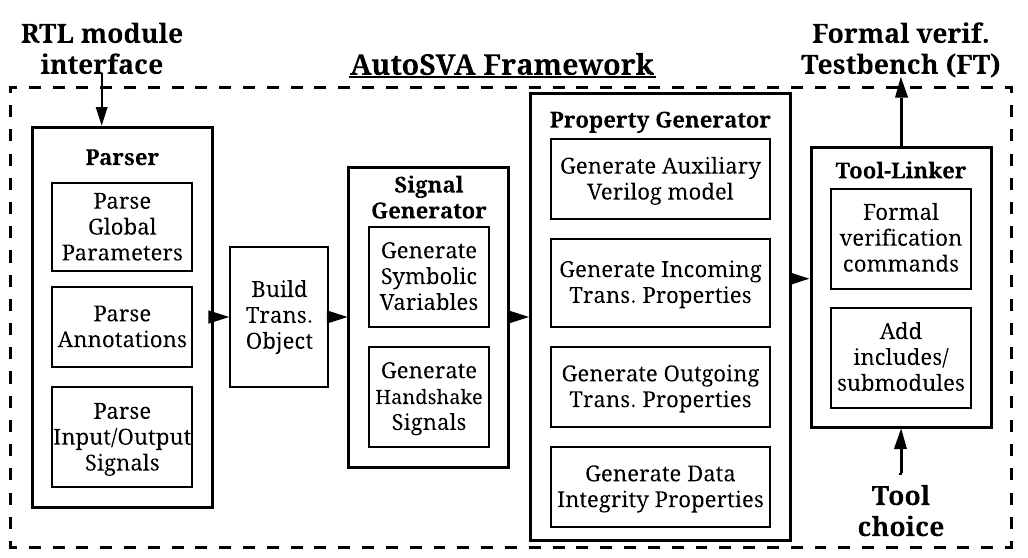}
    \vspace{-4mm}
    \caption{Steps of the \proj framework. It receives an annotated RTL file and the FV tool to target, and it outputs a FT that is ready to be run.}
    \label{fig:autosva}
    \vspace{-4mm}
\end{figure}

\proj is implemented in Python using only standard libraries to provide portability and ease of use. \textbf{\proj{} generates FTs in under a second}. Fig.~\ref{fig:autosva} details its five steps.


\textit{(1)~Parser:} 
\proj parses the signal declaration section of the annotated RTL file to identify global parameters, e.g. cache associativity or queue size, annotations in the \proj{} language, and interface input/output signals. 
Based on the annotations, the parser identifies which pairs of interfaces participate in transactions and creates a mapping from interface pairs ($P$ and $Q$) to a list of their attribute definitions.

\textit{(2)~Transaction Builder:} 
\proj{} builds transaction objects based on interface fields and implication relations identified by the parser. During this process, \proj can detect syntax errors in annotations, e.g. when \textit{transid} or \textit{data} fields are defined in only one of the interfaces of a transaction, or with mismatched data widths.

\textit{(3)~Signal Generator:}
Before generating properties based on transactions, \proj{} generates auxiliary signals, such as symbolics, which are unassigned variables used to build assertions. Symbolic variables are unconstrained, and allow FV tools to explore all their possible values in a single assertion. For example, a single assertion can be used to reason about all lines of a cache if a symbolic signal is used to index the cacheline. \proj{} also generates handshake signals (as conjunctions of  \textit{val} and \textit{ack}) to indicate that a request or response takes place.

\textit{(4)~Property Generator:}
\proj{} creates properties based on the transaction attributes and type (incoming or outgoing). These properties can verify liveness, uniqueness, data integrity, stability, or X-propagation (detailed in Section ~\ref{sec:properties}).
SVA properties are explicitly written in the property file. \proj does not use SVA macros or checkers to provide better readability in case the user wants to explore the properties or a verification engineer wants to extend the FT for functional correctness. The properties are tool-agnostic, and written to be most efficient for FV tools to run, e.g. using symbolic indexes for \textit{transid} tracking. The authors have created this tool based on lessons learnt from prior art~\cite{formal_book, power_assertions,cummings2016} and years of industry and academic experience with FV of RTL. 

\textit{(5)~FV Tool Setup:}
Once the SVA properties are generated, \proj{} links them to the FV tool that the hardware designer selects. 
Furthermore, \proj{} supports linking the FTs of submodules of the DUT, that had already been generated, by using script parameters during \proj{}'s invocation.

\section{Evaluating the \proj Framework\label{sec:evaluation}}

\begin{figure}[t]
    \centering
    \includegraphics[width=\columnwidth]{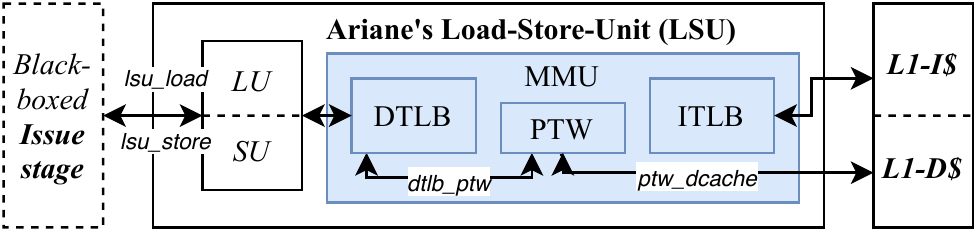}
    \vspace{-3mm}
    \caption{\proj verifies several modules in a hierarchy in Ariane. By testing at the MMU module level (blue box), \proj revealed Bug1.}
    \label{fig:bug1}
\end{figure}

\begin{figure}[t]
    \centering
    \includegraphics[width=0.95\columnwidth]{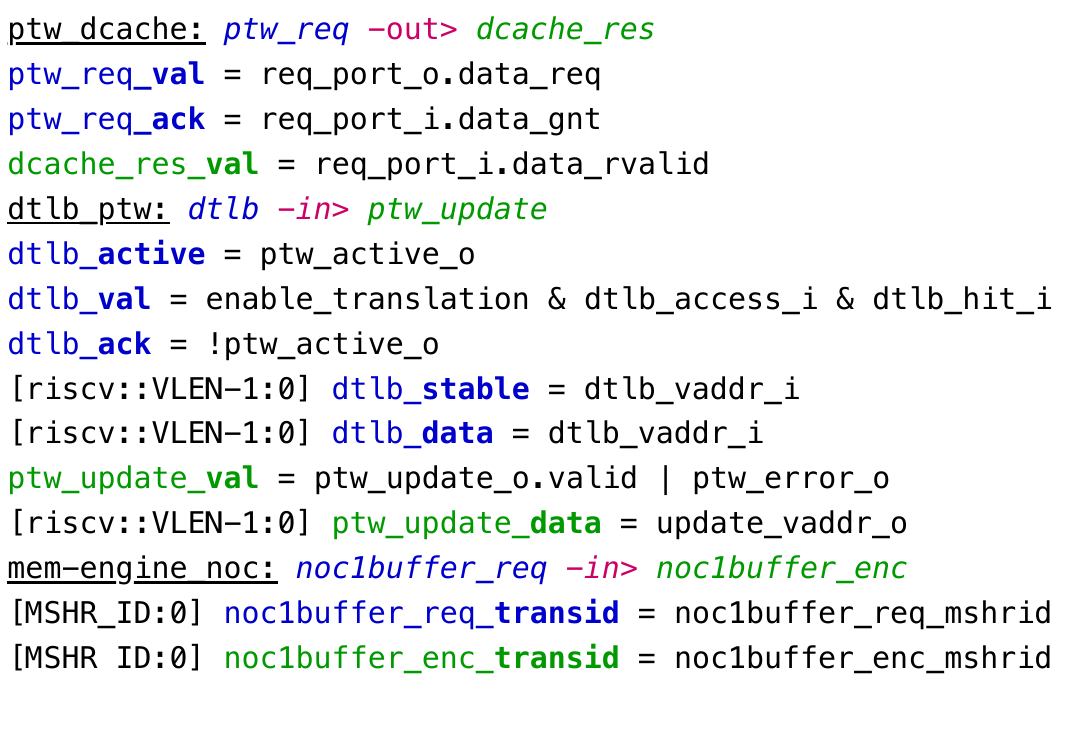}
    \vspace{-3mm}
    \caption{\proj{} annotations to define PTW's outgoing transaction to the data cache (ptw\_dcache) and incoming transaction from the DTLB-miss interface (dtlb\_ptw), and OpenPiton buffer's incoming transaction from \textit{Mem Engine} towards NoC1 encoder (\textit{val} and \textit{ack} attributes match interface names).}
    \label{fig:annotations2}
    \vspace{-3mm}
\end{figure}

We utilize multiple metrics to evaluate \proj: (1)~its \textit{ability to find bugs}, both known (open issues) and new bugs; (2)~the \textit{speed of bug discovery}, based on tool runtime and trace length; (3)~\textit{amount of engineering effort}, measured in time spent writing the transaction annotations; and (4)~\textit{bug-fix confidence}, whether the bug-fix leads to a proof or new CEX.

We study these metrics in mature, taped-out, open-source hardware projects: 64-bit RISC-V Ariane Core~\cite{ariane}
and the OpenPiton manycore framework~\cite{piton+ariane}. We have selected \nomodules RTL modules that are critical for forward-progress and thus require exhaustive testing. Table~\ref{table:rtl_modules} lists these modules as well as the outcome of formally verifying them using testbenches generated by \proj. 
These outcomes consist of proofs and bugs, demonstrating that \proj{} is useful and effective at generating properties and models to verify forward progress.
We also demonstrate \proj{} for early-stage verification by applying it to a new unit, \textit{Mem Engine}, which connects to OpenPiton's NoC by reusing its encode/decoder buffers. 

\proj supports several FV tools, so we elect to perform evaluations using JasperGold 2015.12. Additionally, to check that \proj properties are compatible with system-level simulation, we bind the property files to the in-place testbench using VCS-MX 2018.09.


\textbf{Applying the \proj{} language to RTL modules:} A key component of \proj is its transaction abstraction that is broad enough to apply to most RTL interface styles and specific enough to generate useful properties. 
Fig.~\ref{fig:annotations2} presents a few examples of how \proj{} can be applied to a wide range of interfaces based on common possible scenarios.

\textit{Single Ongoing Transaction:} When there is only one ongoing transaction in a module, it can be modeled simply by not defining the \textit{transid} attribute, which is the case for the \textit{ptw\_dcache} and \textit{dtlb\_ptw} transactions in Fig.~\ref{fig:annotations2}. This principle works for both incoming and outgoing transactions.

\textit{Multiple Outstanding Transactions:} When transactions can be in-flight simultaneously, it can be modeled by annotating the tracking field with \textit{transid}, e.g. \textit{mshrid} for \textit{mem-engine\_noc}. Tracking requests allows \proj reasoning about integrity of \textit{transid} and \textit{data} fields. If requests are not tracked, \proj{} still checks that there are no more responses than requests and that every transaction eventually finishes.

\textit{No Ack signal:} When an interface does not have an \textit{ack} signal but the module cannot always accept new requests, \proj allows defining \textit{ack} by reasoning about other signals. In the case of \textit{dtlb\_ptw}, the \textit{ack} field is defined based on the active signal, that indicates when the PTW is busy. Defining \textit{stable} alongside \textit{ack}
means that \proj will model the payload to remain stable until the request is ack'ed.

\setlength{\tabcolsep}{4pt}
\begin{scriptsize}
\begin{table}[t]
    \centering
    \caption{RTL modules tested with \proj. Ariane modules are indicated with \textit{A}, and OpenPiton with \textit{O}}
    \vspace{-1mm}
    \begin{tabularx}{\columnwidth}{l l}
        \toprule
        \textbf{RTL Module} & \textbf{Result} \\
        \midrule
        A1. Page Table Walker (PTW) & 100\% liveness/safety properties proof\\
        \midrule
        A2. Trans. Look. Buffer (TLB) & 100\% liveness/safety properties proof \\
        \midrule
        A3. Memory Mgmt. Unit (MMU) & Bug found and fixed $-$$>$ 100\% proof \\
        \midrule
        A4. Load Store Unit (LSU) & Hit known bug (issue \#538) \\
        \midrule
        A5. L1-I\$ (write-back) & Hit known bug (issue \#474) \\
        \midrule
        O1. NoC Buffer & Bug found and fixed $-$$>$ 100\% proof \\
        \midrule
        O2. L1.5\$ (private) & NoC Buffer proof, other CEXs \\
        \bottomrule
    \end{tabularx}
    \label{table:rtl_modules}
\vspace{-1mm}
\end{table}
\end{scriptsize}

\textbf{Results:} Table~\ref{table:rtl_modules} presents the \nomodules Ariane and OpenPiton modules that were tested using FTs. \proj generated a total of 236 unique properties based on 110 LoC of annotations. 

First, FTs of Ariane's PTW and TLB resulted in 100\% of the properties being proven at unit-level after 30 minutes of human effort to define the correct transactions.
Next, the MMU FT was set up after 10 minutes of adding a new transaction and reusing the properties of its submodules' FTs.  
These results demonstrate that \proj is quick to use and effective at verifying forward progress in control-critical modules.

Fig.~\ref{fig:bug1} shows the hierarchy of the Ariane modules we have tested. The MMU FT (blue) checks that every request from the LSU eventually receives a response, and that no response occurs without a prior request. Before it uncovered a real bug, \proj found an interesting CEX: an ITLB miss was never filled because the PTW was always busy with DTLB misses, i.e. DTLB has static priority over ITLB. This fairness problem cannot happen in practice since one instruction cannot do many DTLB lookups. Since the trace was quick ($<$1s) and short ($<$4 cycles), it was straightforward to identify the cause of the CEX and add an assumption to remove it.

\textit{Bug1. Ghost Response on MMU:} The next CEX uncovered a bug that was triggered when the MMU receives a misaligned request from the LSU. The MMU responds immediately with a bad alignment response, but the DTLB still misses and the PTW is activated (bad behavior). In the case of a page fault, the MMU generates a second "ghost" response to the LSU, raising an exception. This bug was found by the FV tool in less than a second, producing a 5-cycle trace that allowed us to quickly identify the problem and produce a bug-fix (masking the PTW request with the misaligned signal) with high confidence, as the formal tool found a proof in few seconds for the previously failing assertion. In 5 minutes, the MMU FT proof-rate was 100\%. The Ariane maintainers confirmed the bug and the fix.

\textit{Hitting Known Bugs}: The LSU FT hit (in 1 second) a bug that was recently discovered on a long FPGA run: an ongoing load hits an exception caused by a later load. The Ariane maintainers welcomed a FT where they could validate that the bug-fix solves the problem and does not break anything else. Similarly, the L1-I\$ FT was able to hit a reported bug.

\textit{Bug2. Deadlock in NoC Buffer}: \proj found a deadlock bug in an underdeveloped part of \textit{Mem Engine} that connects to the OpenPiton NoC. 
Since the interfaces mostly matched the \proj{} language, the FT was generated with just 3 lines of code (shown in \textit{mem-engine\_noc} at Fig.~\ref{fig:annotations2}). The first CEX to the liveness assertion revealed a bug that arises from the reuse of the NoC buffer from the L1.5\$ for \textit{Mem Engine}
The buffer assumes that the input does not drive more requests than the number of buffer entries, which is violated in \textit{Mem Engine}. After fixing the bug (adding a "not-full" condition to the \textit{ack} signal), the formal tool resulted in a proof.

Lastly, the FT of OpenPiton's L1.5\$ showed that the condition added to the NoC buffer did not break the properties for its buffer instance.
Other properties, e.g.\ that every cache miss is eventually filled, showed CEXs due to under-constraints in the message types.
\proj provides the FT foundation that the L1.5\$ designer can refine with assumptions to remove spurious CEXs. The testbench can also be extended with more assertions to achieve complete functional verification.

\section{Related Work\label{sec:related_work}}
Early works focused on developing methods for formal verification of RTL correctness~\cite{smc_no_bdds, smc_mcmillan, abv_functional}. 
These include model checking, which relies on SAT solvers or BDDs, and Assertion-Based Verification (ABV), which builds on top of model checking to verify control logic, design interfaces, and data integrity.
The emergence of SVA~\cite{sva_spec} popularized ABV for verification engineers~\cite{formal_book}. More recent work focuses on generating SVA from a higher level language~\cite{rtlcheck, ilang}: 
RTLCheck verifies RTL pipeline implementations against their memory consistency model (MCM) axiomatic specifications;
ILA generates a Verilog model of the design from the ILAng functional specification, and compares it against the RTL implementation~\cite{ilang}.
Although these methods automatically generate SVA, defining high-level models and matching them to RTL signals still require significant effort and knowledge, which may discourage hardware designers from using them.

\vspace{-0.5mm}
\section{Conclusion \label{sec:conclusion}}
This work presents \proj{}, a tool to automatically generate testbenches for unit-level FV. Based on annotations made in the signal declaration section of an RTL module, \proj{} generates liveness and safety properties about control logic to verify forward-progress.
Thus, hardware designers can verify their designs at unit-level without requiring FV expertise and with the minimal effort of writing RTL module interface annotations.
This pays off quickly, as performing FV early can save significant debugging time during system-level simulation and increase designer confidence that the system will not hang.

We have shown that \proj{} is useful and effective with an evaluation on widely used open-source hardware projects. This discovered bugs and provided proofs of \nomodules control-critical RTL modules. \proj generated a total of 236 unique properties (no loops) based on 110 LoC of annotations. 
The generated FTs are extensible and so they are included in the open-source repository of this work\footnote{https://github.com/PrincetonUniversity/AutoSVA}. We envision \proj to become a standard language to define RTL modules' interface expectations, eventually integrated into the SVA specification.



\bibliographystyle{IEEEtranS}
\bibliography{refs}

\end{document}